# Application of flash method in the measurements of interfacial thermal resistance in layered and particulate composite materials

Karol Pietrak[1]*
Tomasz S. Wiśniewski[1]
Michał Kubiś[1]

[1]Institute of Heat Engineering, Warsaw University of Technology

Nowowiejska 21/25, 00-665 Warsaw, Poland

*corresponding author, e-mail: karol.pietrak@itc.pw.edu.pl

**Abstract**

Presented study concerns the possibility of evaluation of interfacial thermal resistance (ITR) between the constituents in composite materials with the use of flash technique. Two variants of such measurement are considered, the first of which is the measurement of ITR between two bonded layers of different materials which had been studied before by various researchers. The second tested measurement method is targeted at determination of ITR in particulate composites with low and moderate filler content based on their effective thermal conductivity. Method of such measurement is proposed and tested on two cases of particle-filled polymer composites. Positive verification results were obtained for polymer/glass composite in which the difference between thermal conductivities of matrix and filler is low. For a polymer filled with aluminum particles the evaluation of average ITR in the samples was impossible as the effective medium models applied in the method strongly underestimated the thermal conductivity of that material. The investigation confirmed the need for more accurate methods of macroscopic thermal properties prediction for composite media with high contrast of thermal conductivities of the constituents. Extended literature study suggests that the method can be applicable to selected classes of engineering materials.

**Keywords:** flash method, thermal contact resistance, interfacial thermal resistance, composite materials, effective medium approximation

## 1 Introduction

From heat transfer perspective, it is convenient to distinguish between two types of solid-solid contacts that occur in technology. The first one will be referred to as non-direct. It occurs for instance between machine parts and structural elements which are fastened together, slide along each other or meet periodically. In such joints, the elements are not connected permanently at the microstructural level. Heat transfer through these contacts is strongly affected by thermal contact resistance (TCR) whose magnitude depends on parameters such as roughness, contact pressure and type of fluid filling the microscopic gaps between the elements [1, 2]. Different type of contact occurs when two materials form a permanent bond, with adhesion and surface matching at microscopic level. It is encountered in composite materials, laminates, thin film solar panels, mechanical joints and printed circuit boards, among many other applications. These types of contacts will be referred to as direct, although they are not free from various imperfections [1, 3]. High quality joints of this type are obtained by careful surface preparation and application of advanced manufacturing techniques, like chemical or physical vapour deposition [4, 5]. To characterize the heat transfer through the boundary between solids remaining in direct contact the notion of interfacial thermal resistance (ITR) is utilized [3, 6]. Apart from thermal contact resistance related to poor mechanical and chemical bonding it includes thermal boundary resistance (TBR) influence which results from the differences in conductive characteristics of contacting media such as velocities and densities of heat carriers [1, 6, 7].

This study concerns the case of direct contact in layered and particulate composite structures. It must be said that measurement of ITR in such connections is highly challenging due to microscopic scale and subtle nature of the phenomenon [3]. For a long time, it was not generally recognized that a thermal resistance would occur between solids remaining in direct contact in ambient temperatures [8]. Effects related with thermal boundary resistance were expected primarily at cryogenic temperatures, and most of older measurement reports covers this temperature range [7]. In 1989 Swarz and Pohl measured TBR of various metals deposited on dielectric substrates in temperatures 1-300 K using electrical resistance thermometry [4]. A similar study was performed in 1993 by Stoner and Maris for temperatures 50-300 K, this time with the aid of transient thermoreflectance technique [9]. These measurements concerned layered structures with highly perfect boundaries in which the interfacial thermal barrier was limited only to the TBR component (no TCR). The values of interface thermal resistance of such bonds are typically of the order of $10^{-8} \div 10^{-7}$ $m^2$ K $W^{-1}$ in room temperature [4, 9]. In many practical applications, like functional composite materials, mechanical contact between the constituent media is imperfect and thus higher values of ITR are obtained in the measurements (~ $10^{-5}$ $m^2$ K $W^{-1}$ [10, 11]).

The presence and significance of interfacial thermal barrier at higher temperatures was noticed in 1980s by researchers devoted to the studies of thermal properties of composite materials [12, 13, 14, 15]. Effective thermal conductivity of these materials exhibited strong positive dependence on temperature, which was attributed to the thermal expansion mismatch effect. It was observed that the difference of thermal expansion coefficients of matrix and inclusions leads to debonding and formation of interfacial gaps whose size depends on temperature. These observations motivated the interest in the phenomenon of interfacial thermal resistance and its influence on the thermal conductivity of composite media [16]. In modern composites the matrix is frequently filled with micro- or nanosized particles of high thermal conductivity (e. g. carbon black, diamond, nanoparticles, graphene). Such materials are very sensitive to the effects of interfacial thermal resistance. For a composite with large particles, the area of interfacial contact per unit volume is low and the effective thermal conductivity is not strongly affected by the ITR. When small filler particles are used, the area of contact per unit volume becomes large, and ITR begins to play a dominant role in the heat transfer. In that case, even adding highly conductive filler to the matrix may result in a decrease of thermal conductivity, and not an increase that is expected [17]. In face of these facts, measurement and prediction of ITR in particulate composites is of great practical interest. Its value can be predicted based on analytical models like Acoustic Mismatch Model [7], and Diffuse Mismatch Model [7], which are valid only in extremely low temperatures [18]. Numerous measurement methods have been proposed for the measurement of ITR between solids remaining in direct contact [3]. They can be used to gain information on ITR in ambient temperatures, but not many of them can be applied directly to particulate composites due to microscopic scale of inclusions [3]. Methods applicable to dispersed composites, like the thermal wave method proposed by Garnier [11], require complicated sample preparation and sophisticated experimental setup. For that reason, many researchers turned to indirect methods of evaluation of filler-matrix ITR, in which filler-matrix bond is recreated in larger scale [19, 20] or with basic cylindrical geometry [10]. The downside of such approaches is the uncertainty of proper recreation of contact pressures, interfacial voids and other defects that are present in the actual composites, which have great influence on the value of ITR.

To avoid these flaws, a method which does not require the recreation of filler-matrix bond and relies on the well-known experimental setup of flash technique [21] is examined in this study. Invention of the flash technique by Parker in 1961 [22] allowed for fast and reliable measurements of thermal diffusivity $D$ and specific heat $c_p$ of solids. Later extensions to the method provided the possibility of measurements in high temperatures [23, 24]. Its application for the evaluation of unknown thermal parameters in two- and three-layered solid structures was also proposed [25-29]. In the most advanced and popular variant [28], the measurement of thermal contact resistance between layers and/or unknown thermal diffusivity of one of the layers is possible. Commercial software, bundled with Netzsch LFA 447 laser flash apparatus used in this study, relies on the approach of Hartmann et al. [28] which is based on an analytical solution of unsteady heat conduction in two-layer sample with interlayer thermal resistance. The success of ITR measurement with the use of that approach depends strongly on the level of signal-to-noise ratio [28] and the ratio of thermal resistances of constituent layers. For accurate measurements, it is advised that the thermal resistance of any of the two layers should not exceed 80% [30]. Due to the availability of commercial software for the evaluation of ITR in layered samples, a set of such samples have been prepared for the investigation, but the routine failed to converge for 4 out of 6 samples. To continue the research, a custom data analysis algorithm was implemented in MATLAB which remained stable for all samples and allowed for the estimation of ITR in polymer/aluminum and polymer/glass bonds.

Second stage of research was aimed at the development of a method that would allow for measurements of ITR in samples of particulate composites with nearly-spherical inclusions. An approach was adopted in which effective thermal conductivity of particulate composite sample is calculated based on the effective medium model and is later used in the numerical solution of direct heat conduction problem. The method uses inverse algorithms to fit the simulated thermal response of the sample to the measured one and obtain the values of unknown filler-matrix ITR, while the remaining parameters of the system are provided. It has to be said that flash method has been used before for thermal conductivity measurements in various types of composites [31-38] but the evaluation of interfacial thermal resistance was not investigated by researchers.

## 2 Methods

### 2.1 Evaluation of ITR in two-layer samples

Rear-face temperature responses of double-layer samples subject to pulse-heating at their front-faces were generated with the use of one-dimensional finite-volume model implemented in MATLAB. These were fitted to the signals measured by LFA 447 instrument, with the ITR treated as free parameter. The elements required in this type of estimation are: the heat diffusion model for 2-layer sample presented in section 2.3 and an inverse method for the calculation of unknown parameter. The Levenberg-Marquardt algorithm [39] was used for data fitting. Prior to fitting, the experimental signals were smoothed with the use of central moving average, it was found however that smoothing had little impact on the results.

### 2.2 Evaluation of ITR in particulate composite samples

The schematic of method for the evaluation of ITR in particulate composites is presented in Fig. 1. This approach is similar to the previous one with three distinct modifications. As before, the direct heat transfer problem is solved numerically with the use of finite volume model described in section 2.3 but this time with only one layer. Thermal conductivity $k_{eff}$ of that layer is calculated from analytical expression proposed by Every et al. [17], often referred to as Bruggeman assymetric model, based on known thermophysical properties of the constituents of composite material and the filler-matrix thermal resistance value supplied by the fitting procedure. The density and specific heat of the layer are obtained beforehand in separate measurements, and treated as known. In the case of particulate samples, the model curve was fitted to the experimental curve based on characteristic time $t_{1/2}$, as described in section 2.5.

For such approach to be valid, conditions should be met which allow to treat heat flow in heterogeneous sample as macroscopically homogeneous. Kerrisk [31, 32] has shown analytically that a composite material can be treated as homogeneous in the laser flash measurement if thickness of the sample is greater than 60 to 70 times the scale of inclusions. This matter was investigated further by Lee and Taylor [33, 34] who examined a wide range of different two-component samples with the use of LFA method and found that accurate results are obtained even if the size of dispersions is relatively large compared to the sample thickness (up to 25% of the sample thickness). All samples investigated in this paper meet that criterion. The biggest size of used filler is 300 μm which constitutes about 19% of the respective sample thickness. For the rest of tested samples the particle size is always less than 5% of the sample thickness.

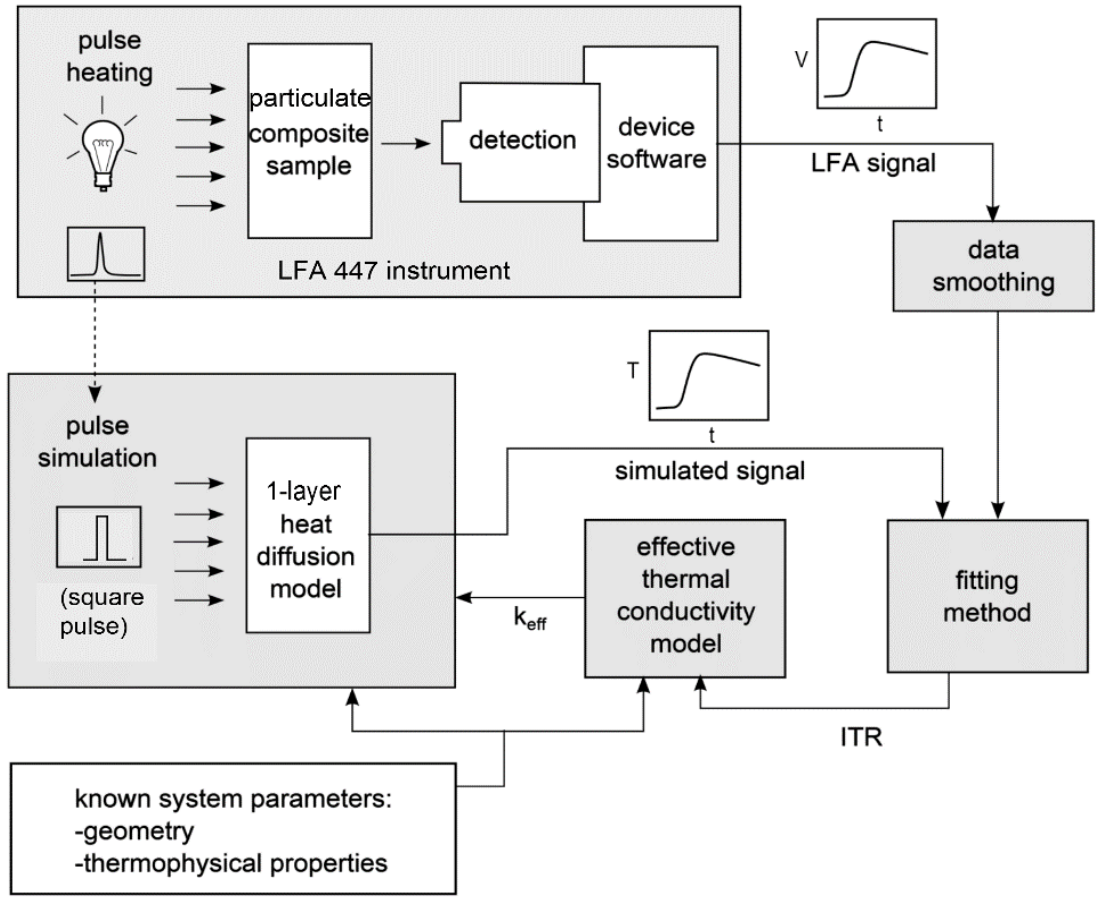

Fig. 1. The schematic of solution method for determination of ITR in particulate composite sample. The LFA signal measured on the sample is smoothed and read by the fitting procedure. Signal from the numerical model is fitted to the experimental curve using the method of bisection. In the numerical simulation of heat conduction, the sample is modeled as homogeneous material of thermal conductivity $k_{eff}$. This thermal conductivity is calculated by the effective thermal conductivity model proposed by Every et al. [17].

**2.3 Numerical model of the forward problem**

The heat conduction within a single- or double-layer sample was treated as one-dimensional due to the geometry of common samples for laser flash analysis, in which the width or horizontal dimension is typically 5 to 10 times greater than height. In such configuration, the lateral area is relatively small compared to the top and bottom areas, so the lateral heat loss is insignificant. The thermophysical properties of both material layers may be treated as constant due to small changes of the average temperature of the sample during experiment, which is under 1 K. Another simplification concerned the area heated by laser or flash lamp pulse. In the one-dimensional model, heating takes place along the whole front face of the sample while in the actual experiment it involved only some limited area around the centre of the surface. The influence of limited heating area is negligible if sensing of the response is carried out directly over the centre of illuminated circle [21]. The non-uniformities of radiation absorption were reduced by the application of graphite coating. For the experimental samples, thin layer of GRAPHIT 33 spray produced by Kontakt Chemie was applied to both front and rear surfaces.

The computational domain representing a two-layer sample was divided into control volumes as shown in Fig. 2. Let us denote by $T_i$ the average temperature of the control volume $i$ of thickness $\Delta y$. Also, let us assign nodes to the centres of the intermediate control volumes, and to the boundaries for the boundary control volumes. Nodes $1,...,l$ belong to layer 1, whereas the nodes $m,...,N$ belong to layer 2. It must be emphasized that there is no interlayer between layer 1 and layer 2. The spacing between these layers is depicted only for the purpose of distinction between the two interfacial nodes.

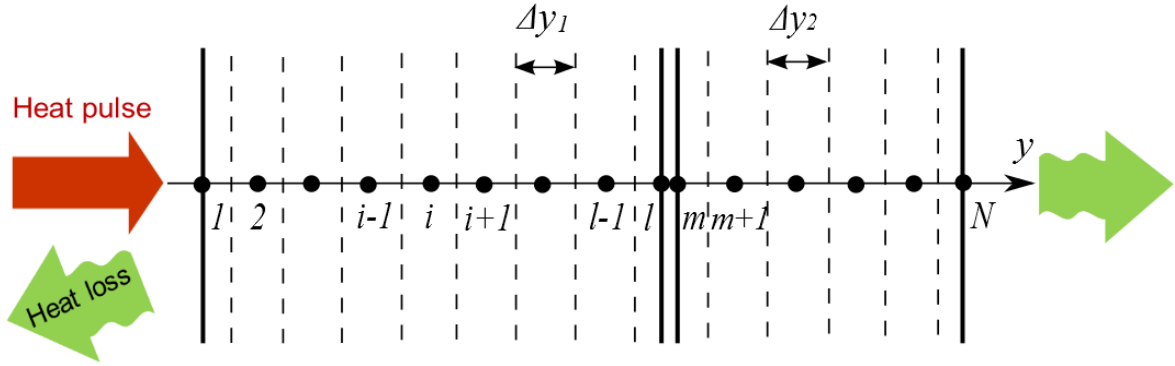

Fig. 2. The schematic of the forward problem and its spatial discretization

For node $i$ ($i = 1,2,...,N$) and layer $j$ ($j=1,2$), the energy balance equation may be written as:

$$\frac{dT_i}{dt} = \frac{q_{i-1} + q_{i+1}}{c_j \rho_j \Delta y_j}, \tag{1}$$

where:

$$q_{i-1} = \frac{k_j(T_{i-1} - T_i)}{\Delta y_j}, \tag{2}$$

$$q_{i+1} = \frac{k_j(T_{i+1} - T_i)}{\Delta y_j}, \tag{3}$$

and $k$ denotes thermal conductivity, $T$ - temperature, $t$ – time, $c$ – specific heat at constant pressure, $\rho$ – density, and $\Delta y$ – spatial step in $y$ direction.

Equation (1) applies for all nodes (control volumes), except for boundary nodes ($1$, $N$), where heat loss is assumed, and two interfacial temperature nodes $l$ and $m$ – one at each side of the interface, where following algebraic constraints on temperatures were assumed:

$$q_l + q_m = 0, \tag{4}$$

$$\frac{T_l - T_m}{R} + q_m = 0, \tag{5}$$

where:

$$q_l = \frac{k_j(T_{l-1} - T_l)}{\Delta y_j}, \tag{6}$$

$$q_m = \frac{k_j(T_{m+1} - T_m)}{\Delta y_j}, \tag{7}$$

and $R$ denotes the unknown interfacial thermal resistance between the layers. The presence of interfacial thermal resistance introduces a discontinuity of temperature field at the interface. The heat fluxes at the boundaries are represented by equations:

$$q_1 = h_1(T_{surr} - T_1) + q_S(t), \tag{8}$$

$$q_N = h_N(T_{surr} - T_N), \tag{9}$$

where $h_1$ and $h_N$ are unknown heat transfer coefficients, $T_{surr}$ is the temperature of surroundings and $q_S(t)$ represents the heat flux related with the heating pulse. It is assumed for simplicity that $h_1 = h_N = h$. The heat transfer coefficient is intended to parametrize the combined heat loss resulting from convection, radiation and conduction between the sample and its metal holder.

In the above modeling, only the spatial coordinate is discretized (see equation (1)). Such approach is known as the Method of Lines (MOL) [40]. For the specific problem formulated above, a system of ordinary differential equations (ODEs) mixed with algebraic equations (for interfacial temperatures) was obtained. This system was solved numerically with the use of *ode15s* solver, which allows for the solution of ODE systems with singular jacobian matrix. The solver is a part of the MATLAB ODE suite, and is described in more detail by Shampine [41].

The stability of the numerical scheme is assured by the inner workings of applied solver [41]. Nevertheless, for an accurate simulation, proper discretization is needed, that considers the timescales and geometry of the modeled phenomenon. The simulation begins with a 2 ms long square heat pulse at the left boundary (see Fig. 2), and this sets the time scale of the problem. The solver adjusts the time step automatically, but initial and maximal time steps must be defined. The initial time step was set to a sufficiently small value of 0.01 ms while the maximal allowed time step $\Delta t$ was set to 0.1 ms. The step of spatial discretization $\Delta y_j$, for a given layer $j$, was selected automatically based on desired Fourier number Fo, according to equation:

$$\Delta y_j = \sqrt{\frac{D_j \, \Delta t}{\text{Fo}}}, \qquad (10)$$

where $D_j$ is the diffusivity of layer $j$. The desired Fourier number was set to 0.5, which is a heuristic approach based on the condition of stability (Fo ≤ 0.5) for fully explicit numerical schemes used for the integration of partial differential equations [42]. Even though the utilized numerical scheme is of different type, this approach allowed to automate the selection of spatial discretization step. The influence of discretization parameters on the obtained solutions was checked. They depended on $\Delta t$, but converged to the same solution for smaller values of the maximal time step. To guarantee solution correctness, sufficiently small $\Delta t$ values were selected for each sample.

**2.4 Initial and boundary conditions for numerical simulations**

The values of thermophysical properties of materials were assumed based on thermal measurements described in section 2.7. The initial temperature of sample and surroundings in simulations was set to zero. The thermal excitation was modeled as a square pulse of duration 2 ms and energy 10 J. Such characteristics were selected in accordance with Hartmann et al. [28].

Determination of heat transfer coefficient $h$ was performed by the Levenberg-Marquardt algorithm in case of layered samples and by a custom fitting procedure in case of particulate composites. The latter task was performed iteratively, with the aid of simulated temperature curves form the presented numerical model. In these simulations, the interfacial thermal resistance parameter was set to zero. The cooling part of the simulated curve was adjusted to the measurement curve until the same slope was reached (see Fig. 3). Heat transfer coefficients obtained with this method for particle-filled polymers were in the range of 40-90 W m$^{-2}$ K$^{-1}$. For layered samples composed of aluminum substrate and polymer coating, very high heat transfer coefficient values were obtained (5500-9000 W m$^{-2}$ K$^{-1}$) which cannot be explained as resulting from sole natural convection. The increased heat loss from these samples occurred due to conduction to the metal sample holder. The area of contact between the sample and the holder was higher than standard because of the non-standard width of the samples. This drawback may be solved using insulation in the future experiments. Sensitivity analysis proved that the heat loss parameter has weak influence on the obtained ITR values (± 4 %), which assures the accuracy of obtained results.

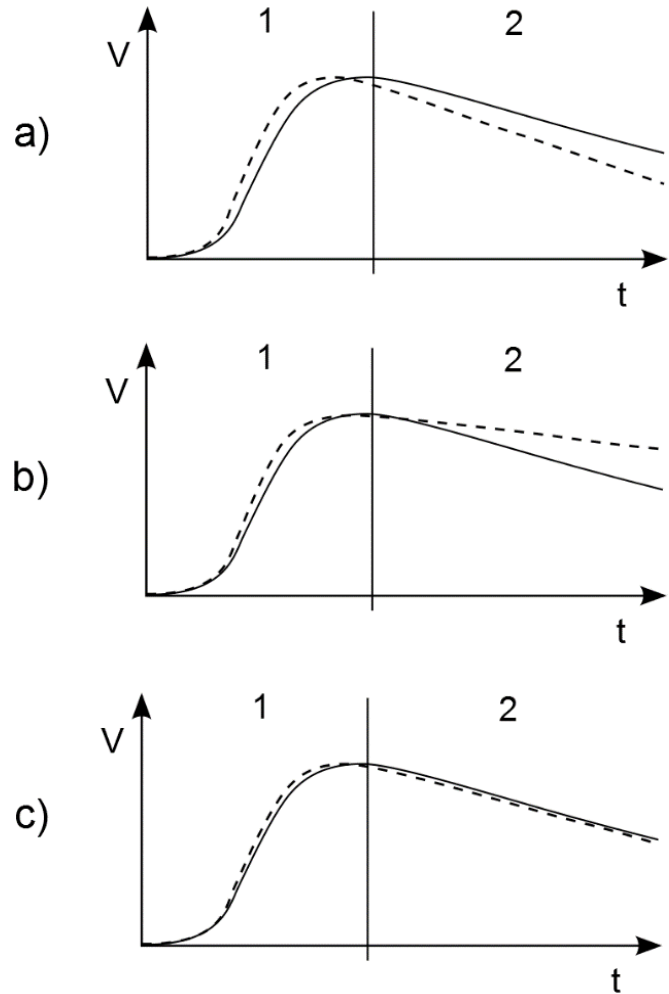

Fig. 3. The influence of heat transfer coefficient value on the shape of LFA curves. In the algorithm of determination of heat transfer coefficient, the signal measured by the flash diffusometer was divided into two parts: 1 – the region of increasing temperature and 2 – the region of cooling down or decreasing temperature. Determination of heat transfer coefficient $h$ was performed iteratively with the assumption of zero ITR. Its value was adjusted until the same slopes of both modeled and measured signals in region 2 were obtained – the case denoted as (c). Case (a) depicts the situation in which $h$ value selected in the model is too high, case (b) – too low. Solid line – experimental curve, dotted line – model, V – voltage, t – time. Please note that the signals are normalized.

**2.5 ITR fitting method for particulate composites**

ITR evaluation method for particulate composites utilized a custom fitting algorithm based on the bisection method [43]. First step concerned the determination of the half-rise time $t_{1/2}$ for the measured LFA signal. This characteristic time constant was originally used for thermal diffusivity measurements by Parker et al. [22] and denotes the time in which the signal reaches half of its maximum. In the next stage, normalized curves from the numerical model were generated iteratively as the values of ITR parameter were varied to minimize the difference between the modeled and experimental signal value at $t_{1/2}$. The method of bisection required the selection of the boundary values of the search interval. The values of $10^{-3}$ and $10^{-7}$ m² K W$^{-1}$ were chosen, which assured that the optimal value is between (see Fig. 4).

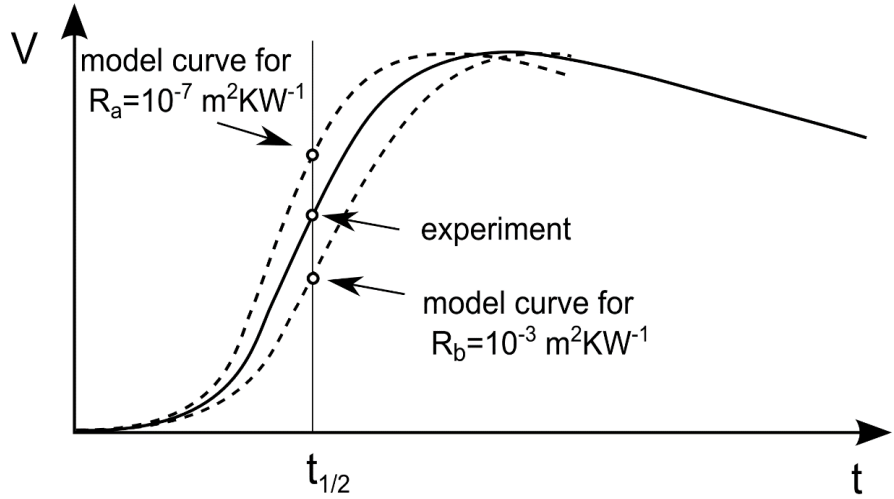

Fig. 4. Dependence of modelled (dotted lines) and experimental (solid line) LFA curves on the ITR parameter. Bisection algorithm was adopted to find the value of ITR that minimizes the difference between the model and experiment at point $t_{1/2}$. $R_a$ and $R_b$ denote the boundaries of the ITR search interval.

**2.6 Effective thermal conductivity models**

The calculation of effective thermal conductivity of particulate composite samples in second tested method (section 2.2) employed assymetric Bruggeman model [17]. The model was chosen after an extensive comparative study aimed at the identification of most accurate and cost effective predictive schemes for effective thermal conductivity of general composite media [16]. This mean-field model assumes random distribution of spherical dispersions and is valid for low-to-moderate volume fractions of filler ($\leq$ 40%). Among the parameters included in the Bruggeman assymetric model are filler volume fraction $\phi$, thermal conductivities of matrix $k_m$ and filler $k_f$, and dimensionless parameter $\alpha$ which describes the effect of inclusion radius $a$ and interfacial thermal resistance $R_{int}$. In that model, the effective thermal conductivity of the composite $k_{eff}$ is given in an implicit form:

$$(1-\phi)^3 = \left(\frac{k_m}{k_{eff}}\right)^{(1+2\alpha)/(1-\alpha)} \left(\frac{k_{eff} - k_f(1-\alpha)}{k_m - k_f(1-\alpha)}\right)^{3/(1-\alpha)} \tag{11}$$

where:

$$\alpha = \frac{R_{int} k_m}{a}. \tag{12}$$

Apart from Bruggeman assymatric model, the Hasselman-Johnson model for spherical inclusions [13] and resistance network model of Yuan and Luo [44] were also considered for thermal conductivity calculations. Contrary to former models, the latter permits modeling of conductive chains and clusters of particles and their impact on effective conductivity. The comparison of predictions of mentioned models for tested composites is shown in section 5.

**2.7 Thermal measurement methods**

Specific heats of specific materials used for fabrication of examined composites were determined in the temperature range 30-100°C, using Perkin-Elmer DSC 7 differential scanning calorimeter. The uncertainty of this measurement is less than 3% [45]. If necessary, the data were extrapolated to lower or higher temperatures (i.e. 25 and 125°C). Densities of polymer matrix and materials present in layered samples were measured with a RADWAG AS-series analytical balance utilizing the underwater weighing technique, with the relative error of ±1.5%. For the fillers, high accuracy density measurement by Micrometrics AccuPyc II 1340 helium pycnometer was conducted with the relative error of ±0.3%. All density measurements were carried out in ambient temperature and the density was assumed constant in the examined temperature range. Thermal diffusivities of the layers in 2-layer samples were obtained with the Netzsch LFA447 diffusometer, by measurements on single-layer samples in temperature range 25-125°C, subject to ± 3% uncertainty, according to the instrument specifications. The thermal model of choice was Cowan model [23] with the pulse correction proposed by Lechner and Hahne [46]. Thermal diffusivities of filler particles are extremely hard to measure due to their microscopic size. Due to that they were assumed same as the diffusivities of glass and aluminum plates of which the layered samples were composed. It must be noted that the error introduced by such assumption is negligible for aluminum-filled polymer, where the contrast of filler and matrix conductivity is as high as 1000 [17]. Moreover, this error is believed to be insignificantly small also for glass-filled polymer, where the conductivity of the matrix has much higher impact on the effective conductivity value.

# 3  Experimental

### 3.1 Preparation of layered samples

Tab. 1 shows precise geometrical data and types of materials used in layered samples. In the table, "top layer" means that the layer was on the detector side and "bottom layer" – the heated side during the flash measurements. The samples were prepared as square plates of substrate material with thin layer of epoxy hardened on the top of it (see Fig. 5 for a SEM cross-section of the epoxy/aluminum sample).

Tab. 1. Geometrical data of tested 2-layer samples

| Sample name | Geometry | Layers/materials | Layer thickness [mm] |
|---|---|---|---|
| Al/Epolam 1 | Square 25 × 25 mm | Epolam2031/2031 (top) | 0.146 |
| | | Aluminum 1050A (bottom) | 1.970 |
| Al/Epolam 2 | Square 25 × 25 mm | Epolam2031/2031 (top) | 0.168 |
| | | Aluminum 1050A (bottom) | 1.970 |
| Glass/Epolam 1 | Square 25 × 25 mm | Epolam2031/2031 (top) | 0.8840 |
| | | Optical glass (bottom) | 0.5260 |
| Glass/Epolam 2 | Square 28 × 28 mm | Epolam2031/2031 (top) | 0.8430 |
| | | Optical glass (bottom) | 0.5510 |

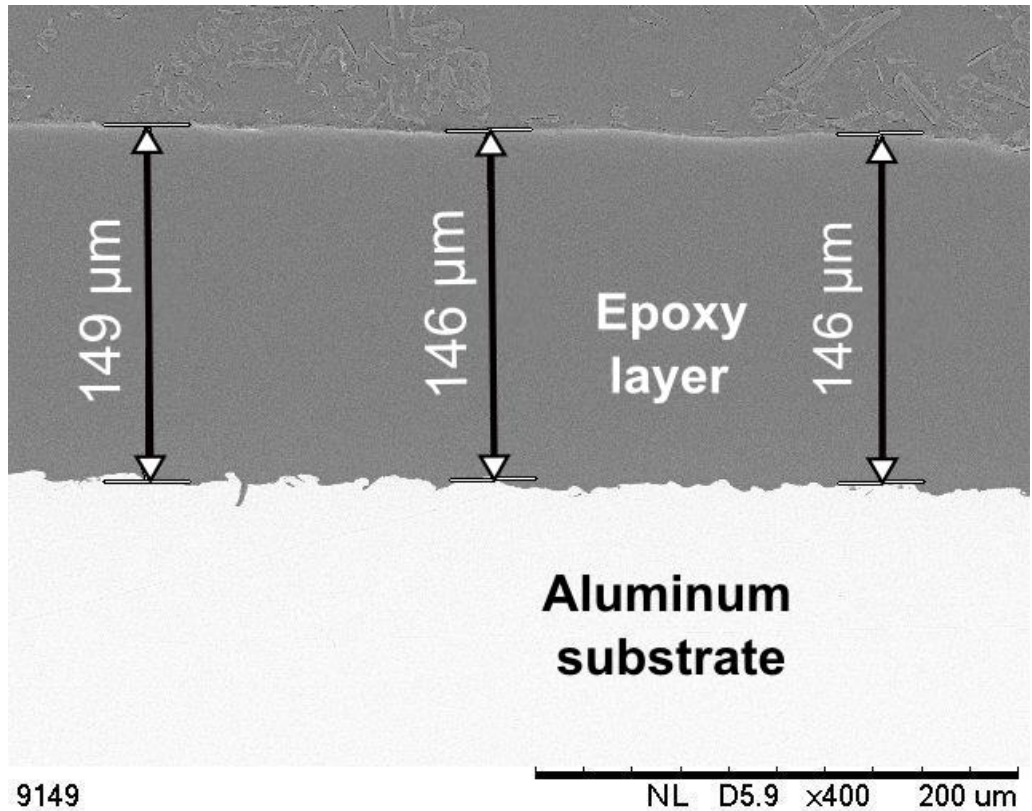

Fig. 5. SEM image of the cross-section fragment of the epoxy/aluminum, showing the results of layer thickness measurement. The waviness of the aluminum surface is also visible

The epoxy mixture used in the experimental samples consisted of *Epolam 2031* resin and *Epolam 2031* hardener prepared by hand mixing with 100:26 mass ratio (reisn/hardener). The aluminum substrates were cut out from the 1050A alloy aluminum profile. Thin optical glass plates were used as substrates in the epoxy/glass samples. The chemical composition of the glass is unknown. Immediately after preparation, the epoxy mixture was poured onto the substrates of known thicknesses. The surface of the substrate plates was in every case cleaned with acetone and polished with P240-grit sandpaper, before applying the epoxy layer, to obtain flatness and good adhesion at the interface.

The samples were further left for 12 hours for initial drying in room conditions. During that phase, most of the air bubbles that had accumulated in the mixture was released and removed from the epoxy layer due to buoyancy. At the next stage, the samples were thermally treated in a furnace for 12 hours. For epoxy/aluminum samples the temperature of thermal treatment was 100°C. For the epoxy/glass layered samples, the 100°C drying was not applied to prevent the glass from breaking associated with thermal expansion. Instead of that, a lower temperature of 50°C have been utilized. After the drying, the epoxy layer was hand-shaped using P240-grit sandpaper to remove excess material, provide flatness, desired thickness and parallelism of the sample planes. Abrasive treatment of the surfaces also enhanced the energy absorption and neutralized reflections which are unwanted during data collection by the infrared sensor.

### 3.2 Preparation of dispersed composite samples

Cylindrical samples of different filler volume fractions and average filler diameters have been manufactured. The detailed data on the composition and geometry of specific samples is given in Tab. 2. The polymer epoxy used as a matrix was the same as in layered samples. One set of samples contained industrial glass microspheres while the second was filled with pyrotechnical atomized aluminum. The particle sizes separation was obtained with the aid of vibrational sieving technique. To obtain samples of desired geometrical characteristics with minimum effort, silicone molds of cylindrical shape were prepared. The epoxy resin was mixed with the hardener by hand, using a beaker and a glass rod. Soon after, filler powder was added and hand-mixed with the resin until the mixture appeared to be well homogenized. It was then cast into the molds for initial 24 h drying in the room temperature. During that phase, most of the air bubbles escaped from the mixture due to buoyancy. At the next stage, the samples were placed for 12 hours in a furnace set to 100°C. After thermal treatment, they were hand-shaped using P100 and P240-grit sandpaper to remove excess material, provide flatness, desired thickness and parallelism of the sample planes. The SEM micrographs of manufactured samples are shown in Fig. 6.

Tab. 2. The geometrical data of manufactured particulate composite samples

| Sample name | Filler volume fraction | Mean filler diameter | Max. filler diameter | Min. filler diameter | Sample Thickness [mm] | Sample Diameter [mm] |
|---|---|---|---|---|---|---|
| Ep-Al 20-40 10% | 0.1 | 30 μm | 40 μm | 20 μm | 1.645 | 12.536 |
| Ep-Al 20-40 20% | 0.2 | 30 μm | 40 μm | 20 μm | 1.879 | 12.556 |
| Ep-Al 20-40 30% | 0.3 | 30 μm | 40 μm | 20 μm | 1.775 | 12.543 |
| Ep-Al 20-40 40% | 0.4 | 30 μm | 40 μm | 20 μm | 1.600 | 13.109 |
| Ep-Glass 40-63 10% | 0.1 | 51.5 μm | 63 μm | 40 μm | 1.594 | 12.32 |
| Ep-Glass 40-63 20% | 0.2 | 51.5 μm | 63 μm | 40 μm | 1.154 | 12.33 |
| Ep-Glass 40-63 30% | 0.3 | 51.5 μm | 63 μm | 40 μm | 1.557 | 12.37 |
| Ep-Glass 40-63 40% | 0.4 | 51.5 μm | 63 μm | 40 μm | 1.964 | 15.74 |
| Ep-Glass 20-40 30% | 0.3 | 30 μm | 40 μm | 20 μm | 1.226 | 12.17 |
| Ep-Glass 02-03 30% | 0.3 | 250 μm | 300 μm | 200 μm | 1.572 | 12.27 |

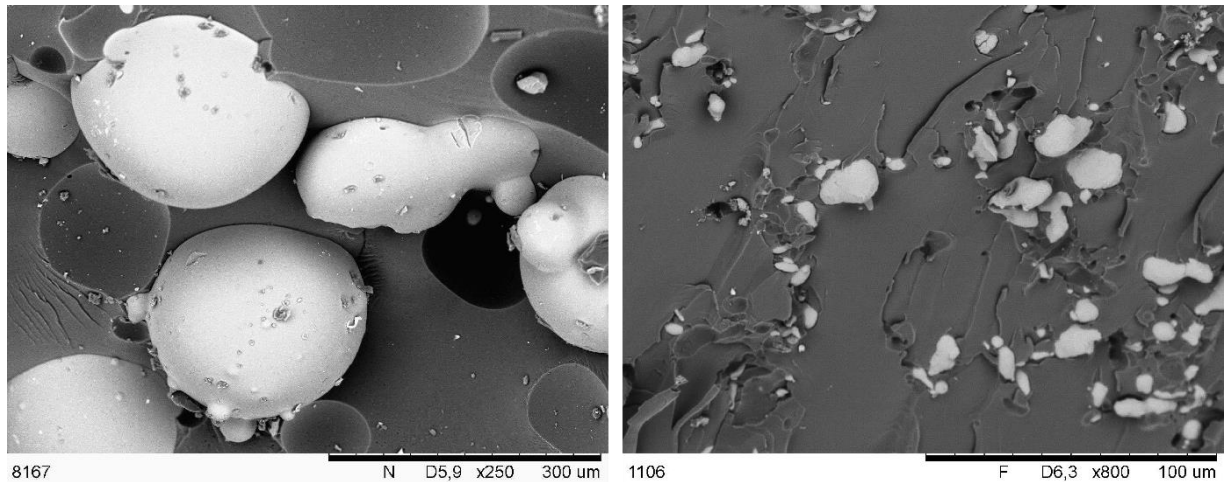

Fig 6. SEM images of tested composites – epoxy filled with glass microspheres (left) and epoxy filled with atomized aluminum particles (right). The mean diameter of glass particles for the epoxy/glass sample is 250 μm, filler volume fraction is 30%. For the aluminum sample the mean filler diameter is 30 μm and volume fraction $\phi = 10\%$.

## 4 Results for layered samples

The interfacial thermal resistance for flat rectangular samples have been determined in the temperature range of 25 to 125°C with the step of 25°C. Three flash measurements were taken at each step and the mean value is shown in the tables as representative. The first case concerned the interface between epoxy resin and aluminum. No signs of layers' separation due to thermal expansion mismatch were visible with naked eye or in SEM micrographs (Fig. 5). Adhesion between two layers was judged as very good as the joint sustained the forces that were exerted during removal of excess material using abrasive techniques. In Fig. 7 it is shown that the ITR variation for the analyzed range of temperatures can be approximated by a straight line, at least up to 100°C. In this region, a constant decrease of ITR was observed. Such character of the curve can be explained by the fact that the sample was thermally treated in 100°C and then cooled down before the measurement was conducted. The epoxy resin hardens completely in 100°C and prior to that it is malleable, so it is reasonable to assume that it adapted itself to the shape of the aluminum substrate at 100°C. When the cooling started, the epoxy layer had already been stiff. Due to the thermal expansion mismatch, microscopic gaps between the two phases grow bigger as the sample cools down. The value of ITR is not examined above 125°C, it can be seen however that for 100-125°C the rate of its decrease is slightly lower than for 0-100°C.

Let us compare the results (Tab. 3) with the literature ones. At 25°C an average value of 4.624 x $10^{-4}$ $m^2KW^{-1}$ is obtained which is close to the result of by Chapelle et al. [10] obtained with hot wire method, where the ITR between polymer matrix and copper wire was estimated to be 3.0 x $10^{-4}$ and 1.6 x $10^{-5}$ $m^2 \cdot K$ $W^{-1}$ for two different wire diameters. Compared to the result obtained for polymer matrix and aluminum fibre by Garnier et al. [11] (ITR = 3.81 x $10^{-5}$ $m^2$ K $W^{-1}$) using thermal wave method, present result is 10 times greater. It can be caused by different geometries and different interfacial conditions existing in a composite with reinforcement.

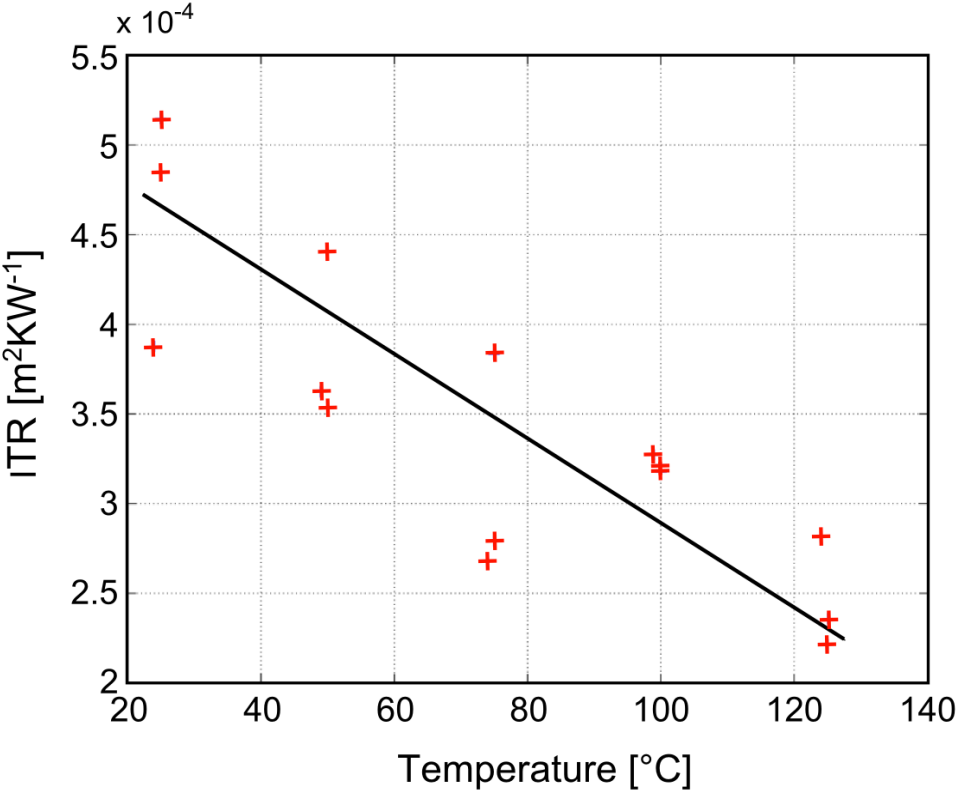

Fig. 7. Interfacial thermal resistance vs. temperature for epoxy-aluminum layered sample. The measurements were carried out in the temperature range of 25-125°C with the step of 25°C. At each temperature, three LFA measurements were taken. Each data point represents the value of interfacial thermal resistance obtained for a single measurement. The black solid line represents a linear regression for the given set of data.

Tab. 3. Mean values of ITR between aluminum and epoxy at different temperatures

| Temperature [°C] | ITR [$m^2KW^{-1}$] | Std. dev [$m^2KW^{-1}$] |
| --- | --- | --- |
| 25 | 4.624 × $10^{-4}$ | 6.635 × $10^{-5}$ |
| 50 | 3.853 × $10^{-4}$ | 4.800 × $10^{-5}$ |
| 75 | 3.102 × $10^{-4}$ | 6.372 × $10^{-5}$ |
| 100 | 3.219 × $10^{-4}$ | 4.676 × $10^{-6}$ |
| 125 | 2.459 × $10^{-4}$ | 3.140 × $10^{-5}$ |

The second type of tested layered samples were epoxy/glass samples and the results for one of them are presented in Fig. 8 and Tab. 4. In the case of these samples, separation of the layers has been observed after the thermal treatment. There were cracks in the glass layer through which air accessed the interfacial gaps. The area of the gaps constituted about 50% of the total area of contact. In the remaining part the adhesion between the glass and epoxy was preserved. Considering these observations, high interfacial resistance was expected. The mean value obtained for room temperature is $8.552 \times 10^{-4}$ m$^2$ K W$^{-1}$.

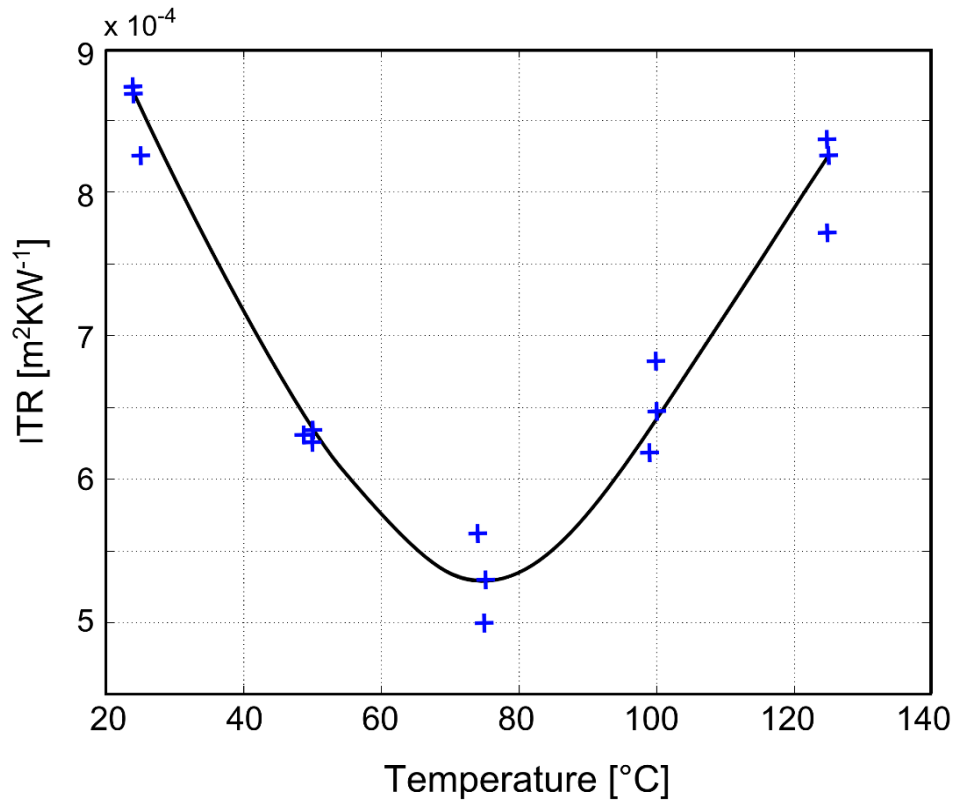

Fig. 8. Interfacial thermal resistance vs temperature for epoxy/glass layered sample measured in the same manner as before (see the description of Fig. 7). One can see that for this type of sample, the character of the ITR variation is different than for the epoxy-aluminum sample. The black solid line represents a spline interpolation of the data.

Tab. 4. Mean values of ITR between glass and epoxy at different temperatures

| Temperature [°C] | ITR [m$^2$KW$^{-1}$] | Std. dev [m$^2$KW$^{-1}$] |
|---:|---:|---:|
| 25 | $8.552 \times 10^{-4}$ | $2.570 \times 10^{-5}$ |
| 50 | $6.300 \times 10^{-4}$ | $4.120 \times 10^{-6}$ |
| 75 | $5.304 \times 10^{-4}$ | $3.141 \times 10^{-5}$ |
| 100 | $6.489 \times 10^{-4}$ | $3.150 \times 10^{-5}$ |
| 125 | $8.116 \times 10^{-4}$ | $3.501 \times 10^{-5}$ |

In Fig. 8 the character of ITR variation with temperature is different than for the epoxy/aluminum sample (Fig. 7). This can be caused by the fact that epoxy/glass samples were subject to different thermal treatment prior to the measurements. In case of these samples, 50°C is the maximal temperature that was reached. It is logical to expect that final hardening of the resin occurred in that temperature and before this event, the resin was plastic (analogically as for epoxy/aluminum samples, but this time in a lower temperature). The epoxy layer adapted its shape to the glass substrate at 50°C and become stiff which resulted in a thermal expansion mismatch that occurred between the layers during cooling down to the room temperature. The minimal ITR is reached for 75°C measurement which is higher than the maximal temperature of the thermal treatment, but for higher temperatures the ITR increases, which suggests that the interfacial gaps become bigger or some other phenomenon contributing to the thermal resistance becomes more pronounced. Nevertheless, this behavior seems to relate to the specifics of thermal treatment that is a part of the manufacturing process.

## 5 Results for particulate composite samples

The comparison of predictions of different effective thermal conductivity models selected based on the literature study was performed before selection of the model that was employed in the inverse calculations of ITR. Results of thermal conductivity measurements for manufactured epoxy/glass composites are shown in Fig. 9, accompanied by the predictions of mentioned models. Thermal conductivity of samples was determined indirectly, based on determinations of thermal diffusivities, specific heats and densities, as described in section 2.7. Both analytical models give similar results, whereas the numerical resistance network model of Yuan and Luo predicts even lower effective conductivity despite the incorporation of percolation effects. If the resistance network model is considered, the interface thermal resistance in the experimental samples is negligibly small, which excluded this model form the ITR calculations. In face of such outcome, and very close results obtained from analytical models, it was decided to calculate ITR using the model proposed by Every et al. considering better behavior of Bruggeman type approximations near the percolation threshold [16].

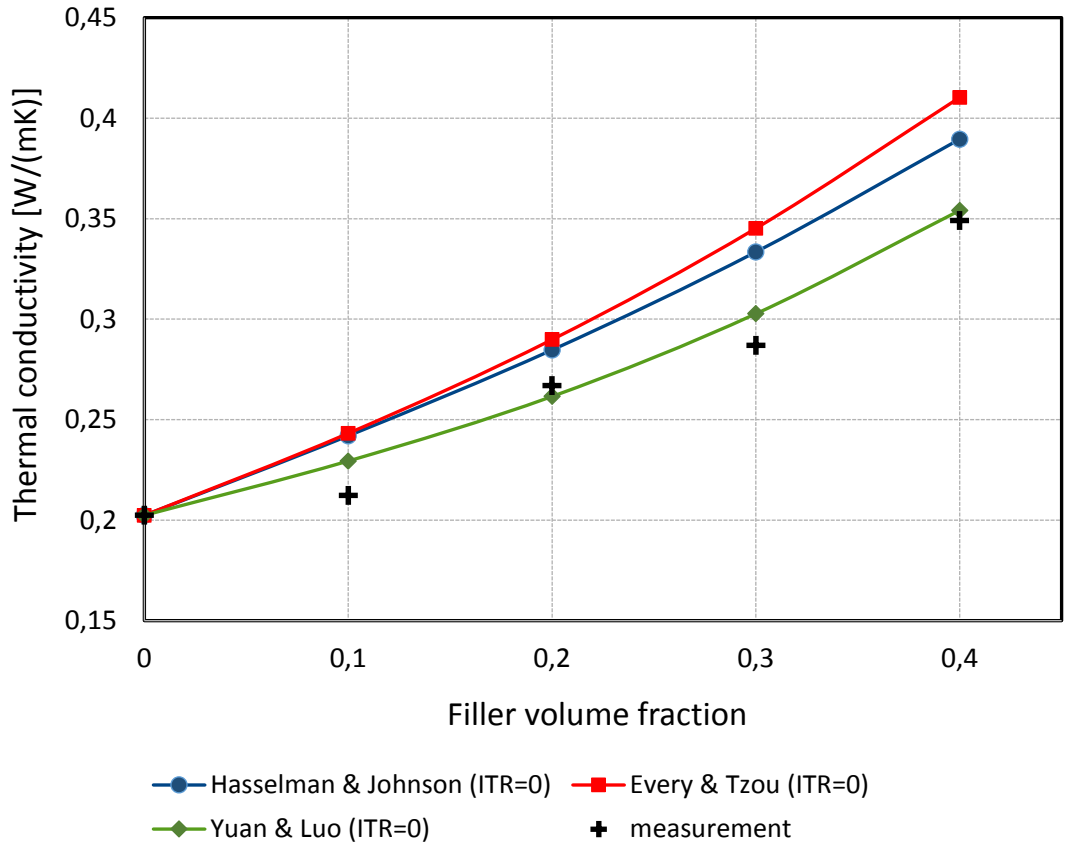

Fig. 9. The effective thermal conductivity of epoxy/glass particulate composite vs. filler volume fraction. Comparison of models with experiment for filler volume fractions up to 40%. The average diameter of glass particles is 51.5 μm.

Interfacial thermal resistance was evaluated for all manufactured samples at different mean sample temperatures from 25 to 100°C with the step of 25°C. Three measurements were taken on given sample per each temperature. The experimental data were analyzed using presented algorithms, which allowed for some preliminary comparisons regarding the influence of temperature, mean diameter of inclusions and their volume fraction on the filler-matrix ITR in the composite. ITR values obtained for samples with different content of glass filler (mean diameter 51.5 μm) in different temperatures are shown in Fig 10 and Table 5. The result for the sample containing 10% of filler is displaced strongly from the remaining points, which suggests that the sample is not representative and probably subject to some manufacturing error. For other samples, the evaluation yielded very similar values, which suggests that ITR calculated with the proposed method does not depend on the percentage of filler for the tested range of filler volume fractions. No strong dependence of ITR on the mean sample temperature can be found within the examined temperature range, except for the anomalous 10% sample.

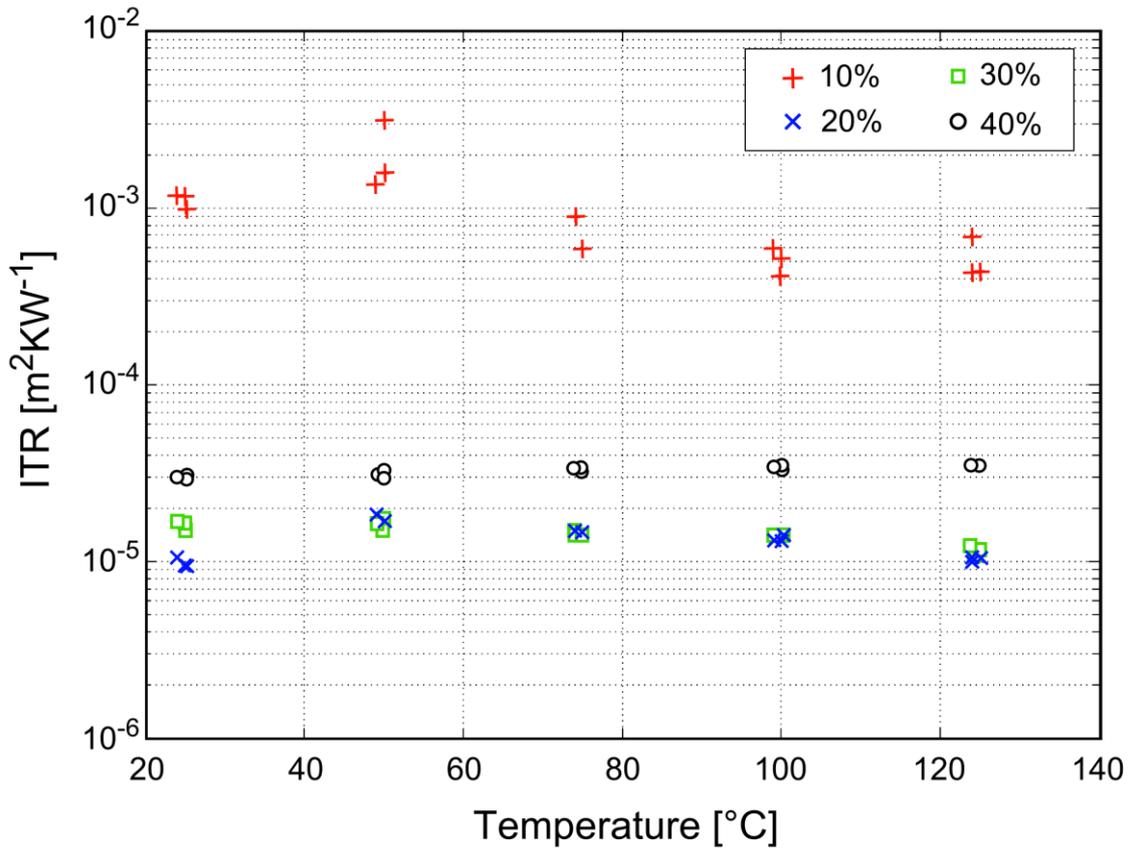

Fig 10. Interfacial thermal resistance vs. temperature for different volume fractions (10%, 20%, 30% and 40%) of glass inclusions, of diameter 51.5 μm, embedded in epoxy matrix. The numerical values of interfacial thermal resistance corresponding to this graph are grouped in Table 5.

Tab. 5. Values of ITR in epoxy/glass composite obtained for different volume fractions of filler (average of 3 measurements). The mean diameter of filler particles is 51.5 μm.

| Temperature [°C] | ITR [$10^{-5}$ m$^2$ K W$^{-1}$] | | | |
|---|---|---|---|---|
| | 10% filler | 20% filler | 30% filler | 40% filler |
| 25 | 110.8 | 0.976 | 2.955 | 1.557 |
| 50 | 201.9 | 1.763 | 3.134 | 1.647 |
| 75 | 68.46 | 1.479 | 3.354 | 1.441 |
| 100 | 50.56 | 1.363 | 3.451 | 1.415 |
| 125 | 51.78 | 1.015 | 3.459 | 1.209 |

Another comparison encompassed samples with the same volume fraction of inclusions (30%) but different mean diameters (30, 51.5 and 250 μm). The plot of ITR measurement results for these samples is shown in Fig 11. For bigger inclusions, higher ITR evaluation error was expected due to the violation of the condition of homogeneity. In Tab. 6 one can see that result for the largest tested filler (250 μm, which is 16% of the sample thickness) is indeed displaced towards higher values of ITR, while the results for smaller inclusions (30 and 51.5 μm, which are less than 5% of the sample thickness) are similar. The tendency of ITR to increase with increasing filler size can be observed, and there is no strong dependence on temperature.

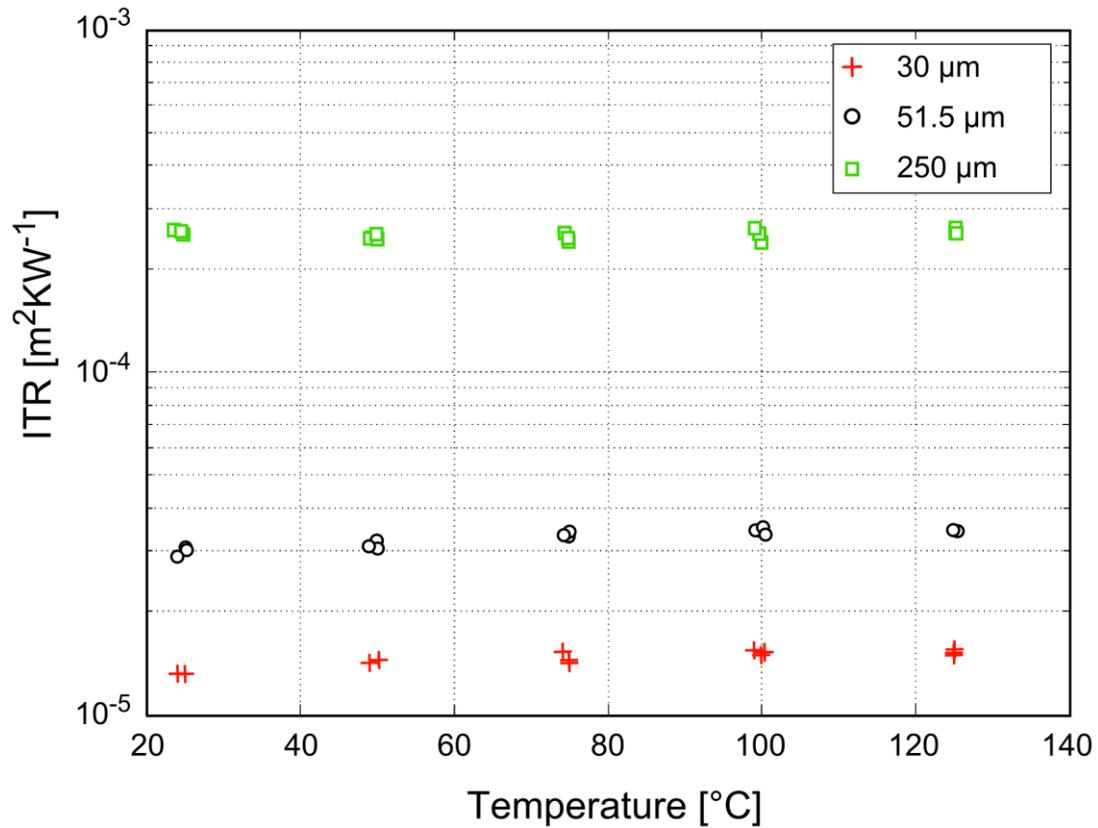

Fig 11. Interfacial thermal resistance vs temperature for different average diameters (30, 51.5 and 250 μm) of glass inclusions embedded in polymer matrix. The volume fraction of the filler is 30% in every case. The numerical values of interfacial thermal resistance corresponding to this graph are grouped in Table 6.

Tab. 6. Values of ITR in epoxy/glass composite obtained for different volume fractions of filler. The mean diameter of filler particles is 51.5 μm.

| Temperature [°C] | ITR [$10^{-5}$ m$^2$ K W$^{-1}$] | | |
|---|---|---|---|
| | 30 μm | 51.5 μm | 250 μm |
| 25 | 1.320 | 2.955 | 25.41 |
| 50 | 1.434 | 3.134 | 25.19 |
| 75 | 1.458 | 3.354 | 24.54 |
| 100 | 1.515 | 3.451 | 25.12 |
| 125 | 1.515 | 3.459 | 25.78 |

Let us now focus on the epoxy/aluminum composite for which an analogical analysis was attempted. For that material, experimental results are well above the models' predictions even if the unrealistic condition of negligible ITR is assumed (see Fig. 12). All models predict that the presence of ITR equal to that measured for layered samples (4.62 x $10^{-4}$ $m^2KW^{-1}$) should cause the decrease of thermal conductivity of the composite to values below the conductivity of the matrix. The reality is very different – experimental thermal conductivity of the composite is approximately 6 times higher than that of the pure for the filler volume fraction of 40%. Such experimental result is not anomalous as Lin et al. [47] reported almost identical thermal conductivity improvement for an aluminum-filled polymer. Therefore, it must be the models that are not adequate for given composite, hypothetically due to the omission of particle shape or clustering influence.

Model proposed by Duan et al. [48, 49] allows for macroscopic thermal conductivity predictions concerning composites with spheroidal inclusions and interfacial thermal barrier. Predictions done with this model have shown that for polymer composites with highly conductive filler and insulating matrix, effective thermal conductivity depends strongly on both the aspect ratio of particles and magnitude of ITR. The aspect ratios of applied aluminum particles are hard to determine due to their irregular geometry. The analysis of SEM images (Fig. 6 shows one of them) proved that the particles are slightly elongated, with aspect ratios of order of 3 to 4. In Fig. 12 one can see the predictions of Duan et al. model for prolate spheroidal inclusions of aspect ratio $r = 3.6$, with and without the interfacial thermal barrier. Good match between the experimental data and model was obtained when ITR is entirely neglected. Precise determination of ITR would require a medium with inclusions of well-defined shape, so that the uncertainty of particle aspect ratio parameter is reduced. In conclusion, the exact values of ITR were not retrieved in case of the epoxy/aluminum particulate samples, as the ITR seemed to be of negligible magnitude, under adopted assumptions. Further investigations of ITR in these samples shall be done in the future.

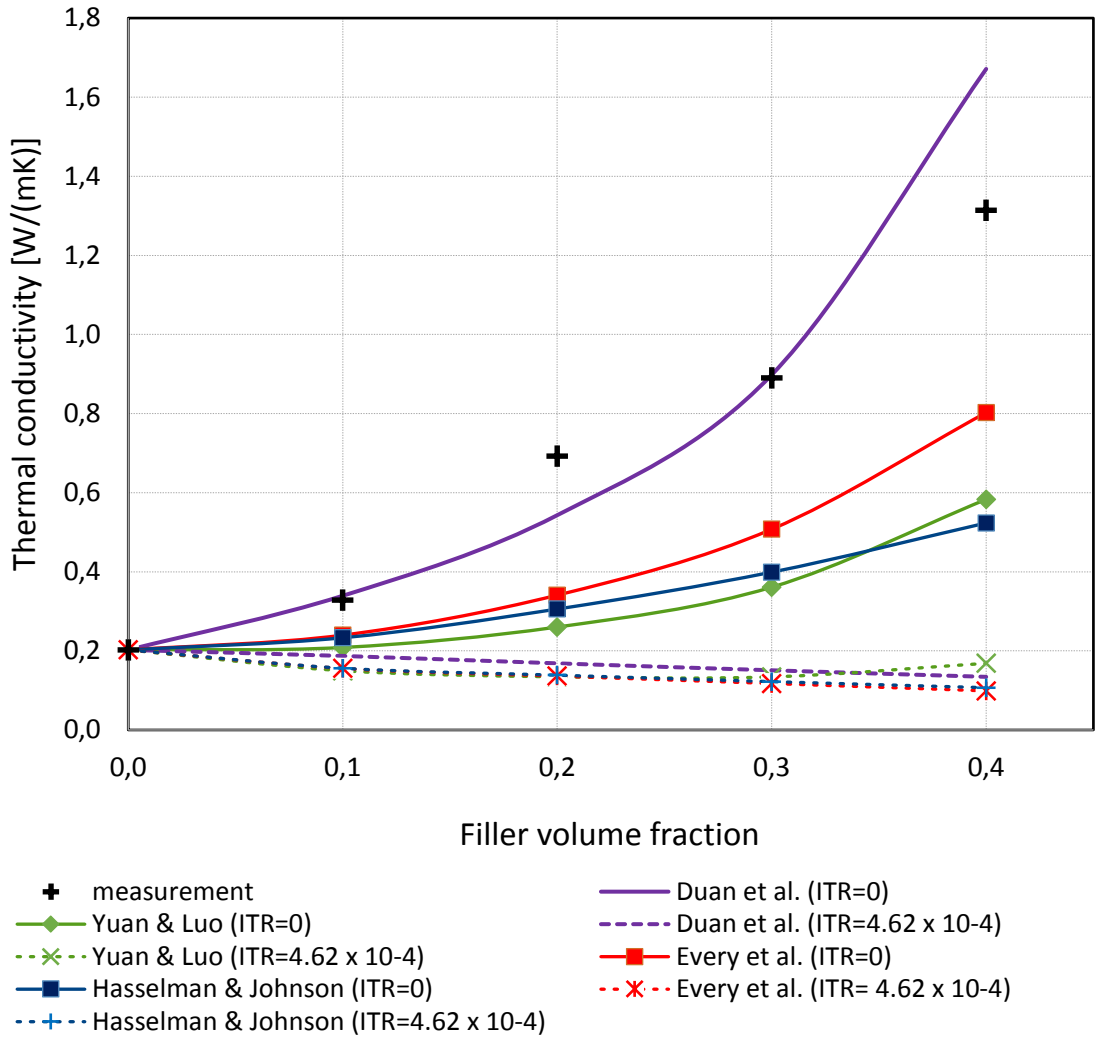

Fig. 12. The effective thermal conductivity of epoxy/aluminum particulate composite vs. filler volume fraction. Comparison of models with experiment for filler volume fractions up to 40%. Average diameter of aluminum particles is 30 μm. The predictions of Duan et al. model [48,49] were fitted to experiment under the assumption of fixed equatorial radius of the spheroids (15 μm) and variable polar length. The best match was obtained for zero ITR and particle aspect ratio $r = 3.6$

# 6  Discussion

The measurement of presented type contained relatively many input variables and assumptions that contributed to the uncertainty of results. Let us consider the particulate samples first. The uncertainty of determination of effective thermal conductivities for these samples by the method described in 2.7 was estimated as ±5 %, considering the contributing errors. Such uncertainty corresponds roughly to one order of magnitude uncertainty of estimated ITR values. Nevertheless, the main source of uncertainty is the applied effective medium model. Depending on the utilized averaging technique, models of this type yield different predictions for the same input parameters and geometries (see e.g. [16] for comparisons).

In case of layered samples, possible errors may result from heat transfer model, measurements of thermophysical properties, layer thickness determination and the quality of measurement-model fitting. Temperature curves generated by the heat transfer model agreed with the curves obtained via the pulse-corrected Cowan model, which proved that the heat transfer modeling is correct. The uncertainty associated with data fitting was calculated in accordance with the statistical method advised by Ozisik and Orlande [39]. In this method, the standard deviation of the *j-th* parameter estimated with Levenberg-Marquardt algorithm is given as [39]:

$$\sigma_{Pj} = \sigma \sqrt{[J^T J]_{jj}^{-1}}, \tag{13}$$

where $\sigma$ is the standard deviation of the experimental data and $J$ is the sensitivity matrix, as defined in [39]. The confidence interval for 99% confidence level of the estimated parameter $P$ can be determined as:

$$\hat{P} - 2.576\, \sigma_P \leq P \leq \hat{P} + 2.576\, \sigma_P. \tag{14}$$

These estimations gave low value of relative ITR uncertainty due to the data fitting, which is ±0.4%.

Despite the uncertainties related with ITR evaluations, the method used in this paper in case of particulate samples has the advantage of utilizing the classical flash technique setup encountered in many laboratories. Commercial instruments of this type with appropriate software allow for fast and automated measurements. Secondly, the method can be easily modified to allow measurements of the filler-matrix ITR in various types of composites simply by switching between different effective thermal conductivity models. At the same time, it was found that the accuracy of predictive models for macroscopic thermal conductivity of heterogeneous materials is the limiting factor for the method. In this and numerous other studies [47, 50, 51] it has been demonstrated that effective medium approximations of types considered in this paper underpredict thermal conductivity of composites with high contrast between thermal conductivities of matrix and inclusions which makes the calculation of filler-matrix ITR impossible.

There are also works that confirm the usefulness of effective medium models in various cases of engineering materials. Stránský et al. [52] showed that evaluations obtained with the use of Mori-Tanaka approach (leading to expression identical with the Hasselman-Johnson model in case of equisized spherical inclusions) agree with experiment for a random dispersion of copper particles in the epoxy matrix (a case very similar to examined here) and as well as for a porous Al/SiC composite. In both case studies presented by Stránský et al. the presence of ITR is considered. Yuan and Luo [44] successfully applied the resistance network approach in the prediction of thermal conductivity of silicone/phosphor composites which are widely used in light emitting diodes (LEDs) packaging. Their model results matched to the experimental data within ±6% in the range of the volume fractions from 3.8% to 25%. Devpura et al. [53, 54] reported validity of a very similar unit cell model in the predictions of thermal conductivity of thermal interface materials employed in flip-chip technology, especially in the higher range of filler volume fractions. Based on those studies it can be inferred that the method presented in this study has the potential of application in selected areas of technology. Further work shall be concentrated on verification of the method with other types of composites and identification of appropriate effective conductivity models, especially for composites with high conductivity contrast.

# 7 Conclusions

In this paper, a method for determination of interfacial thermal resistance in dispersed composite materials that combines the classical experimental setup of flash technique and predictive models for effective thermal conductivity is proposed and its verification results are shown for two cases of composite materials. Prior to the measurements on dispersed composites, the ITR between considered materials was measured in a two-layer configuration, and its dependence on the temperature of thermal treatment was observed. The measurements on dispersed composites revealed that the limiting factor for the applicability of proposed method is the accuracy of predictive schemes for effective thermal conductivity. Presented study, as well as literature review, suggest that there is great need for accurate predictive models for composites with high contrast of conductivities of components, as the classical effective medium approximations often fail in this area. Nevertheless, it was shown that interfacial thermal resistance value in dispersed epoxy/glass composite can be evaluated using flash method, and the usability of this technique for other engineering materials was also discussed based on the available literature.

# 8 Acknowledgements

The support for this research was received from the European Social Fund within the Operational Program Human Capital, project No. POKL.04.01.01-00-061/10 "*Didactic Development Programme of the Faculty of Power and Aeronautical Engineering of the Warsaw University of Technology*". This work was also supported by the Faculty of Power and Aeronautical Engineering of the Warsaw University of Technology in the form of a Dean's Grant.